# Spin Hall utilizing U(1) and SU(2) gauge fields

S. G. Tan,[1] M. B. A. Jalil,[2] Xiong-Jun Liu,[3] and T. Fujita[1,2]

[1]Data Storage Institute, DSI Building, 5 Engineering Drive 1, (Off Kent Ridge Crescent, National University of Singapore) Singapore 117608

[2]Information Storage Materials Laboratory, Electrical and Computer Engineering Department, National University of Singapore, 4 Engineering Drive 3, Singapore 117576

[3]Department of Physics, National University of Singapore, 2 Science Drive 3, Singapore 117542.

We propose a two-dimensional electron gas (2DEG) system in which an external magnetic (B) field with a small chirality is applied to provide a topological $U(1) \otimes U(1)$ gauge field that separates conduction electrons of opposite spins in the transverse direction. Additionally, the vertical electric (E) field in the 2DEG, together with spin-orbit coupling, produces a SU(2) gauge field which reinforces or opposes the effect of the topological gauge. The system thus provides a tunable spin Hall effect, where an applied gate voltage on the 2DEG can be used to modulate the transverse spin current. As this method leads to the enhancement or cancellation of intrinsic spin Hall, it naturally distinguishes the extrinsic from the intrinsic effect.

In recent years, there have been various spintronic propositions with respect to device functionality, foremost of which is the Datta-Das [1] transistor that utilizes the Rashba [2,3,4] spin-orbit (SO) effects to induce spin precession across the two-dimensional electron gas (2DEG) conduction channel. Subsequent experiments [5,6] have confirmed the working principles of such devices, but failed to observe a large conductance modulation. In parallel to these developments, Majumdar *et al.* and others [7,8,9] have shown the theoretical possibility of using external delta ($\delta$) magnetic fields to induce spin polarization. However, such devices are difficult to implement because they require spatially concentrated magnetic fields in order to approximate the $\delta$-function distribution.

Recently, attention has shifted to utilizing the SO coupling to generate a pure spin transport in the transverse direction of the 2DEG, i.e. the spin Hall effect. It has been shown theoretically [10,11] that in the ballistic limits, a pure spin Hall current can be achieved. However, Inoue *et al.* [12,13] showed that the spin Hall conductivity vanishes when vertex corrections are introduced to model the effects of impurity scattering. The present consensus, however, is that such a complete suppression of the spin Hall effect is not a general phenomenon. For instance, in a two-dimensional-hole gas [14], the spin Hall effect persists in the presence of impurities. The origin of the spin Hall effect in the ballistic regime due to the presence of SO (e.g. Rashba) interactions may be understood in terms of the non-abelian SU(2) gauge which gives rise to a transverse force [10] on conduction electrons. The geometry of the 2DEG system is such that to achieve a significant spin Hall current, we require i) spins polarized in the vertical direction and ii) a strong SU(2) gauge field, resulting e.g. from the large electric *E*-field perpendicular to the 2DEG plane. However, the requisite vertically spin-polarized state is not an eigenstate of the system, since the vertical *E*-field produces a relativistic magnetic field in the in-plane direction. This leads to two possible effects, both detrimental to spin Hall: i) the large *E*-field will hasten the relaxation of the initial vertical spins to the in-plane direction, thereby suppressing the SU(2) transverse force, or alternatively, ii) for channel lengths shorter than the spin coherence length, the spin vector will precess about the in-plane relativistic magnetic field, causing a Zitterbewegung-like motion and resulting in zero net transverse spin current. Thus to prevent either the spin relaxation or precession of the vertical spin state, one needs a very short device channel and minimal scattering, making the realization of spin Hall in 2DEGs difficult to implement.

Based on the above discussions, we thus propose a 2DEG system which removes the competition between the SU(2) force and the spin relaxation/precession due to the strong vertical $E$ fields in the 2DEG, and also provide insights into possible experimental implementation. The device utilizes an external magnetic field $B$ and SO coupling. The $B$ fields are applied by means of ferromagnetic gate stripes [15,16] deposited on top of a high-electron-mobile-transistor (HEMT) heterostructure device, as shown in Fig. 1.

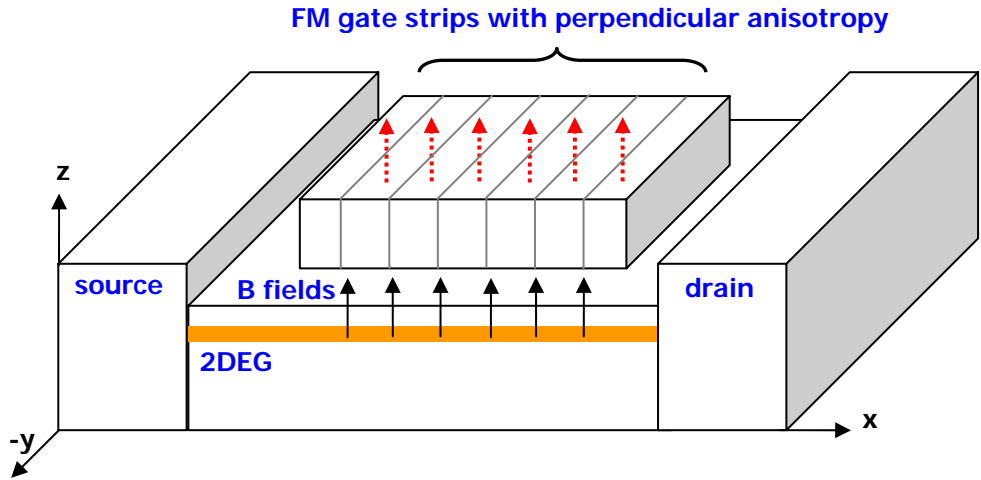

FIG. 1 (color online). Schematic illustration of a lateral device with ferromagnetic gates that utilize the gauge fields to realize the spin Hall effect. The dotted arrows represent the gate magnetization. The schematically vertical B fields represent a two-dimensional distribution of B fields with a small chiral angle as shown in Fig.2.

The $B$ field applied here should be sufficiently strong relative to the in-plane relativistic magnetic field so as to ensure that the two spin eigenstates point in the out-of-plane direction, ideally close to parallel and antiparallel to the $z$-axis. However, we assume the $B$ field has a much weaker influence on spin polarization of current across the device, i.e. we assume a weak polarization in which the conduction electrons are almost evenly split between the two eigenstates. A spatial non-uniformity is superimposed onto the uniform vertical $B$ field in order to introduce a finite chirality (see Fig. 2b). Thus, the applied $B$ field serves a two-fold purpose: 1) to counteract suppression of the SU(2) force due to spin relaxation to the in-plane direction, and 2) to provide a topological spin Hall effect, arising from the spatial non-uniformity of the field. Thus, spin separation is achieved by virtue of the SU(2) gauge, as well as the U(1)$\otimes$U(1) topological gauge [17,18,19] related to the chirality of the external $B$ field. Transistor action can be achieved via the quantum mechanical force due to the former SU(2) gauge, which can be tuned via a gate bias. Figure 2 shows the schematic illustration of the various forces. The quantum Hall effect [20,21] that might arise from the application of external or the effective magnetic fields will not be discussed here.

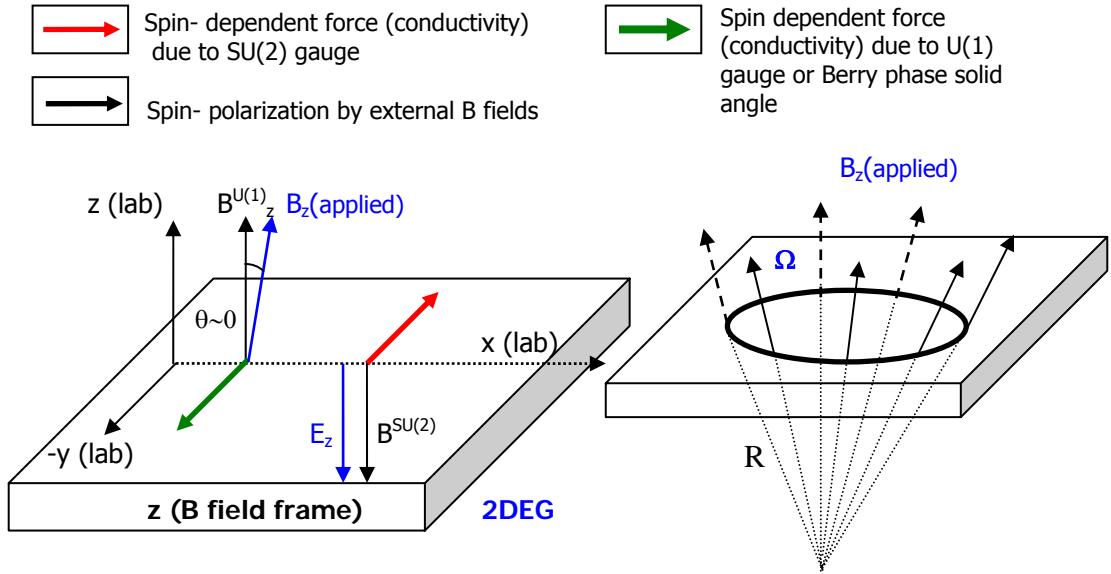

FIG. 2 (color online). (a) Schematic illustration of the forces in opposite directions with proper setting of the field directions. The spin Hall current can be modulated/reversed by tuning and reversing the $E_z$ field. (b) Schematic illustration of applied field distribution with net chirality characterized by solid angle $\Omega$.

To provide a theoretical description of the device, we write the Hamiltonian of the system as

$$H = \frac{p_x^2}{2m} + \frac{(p_y + eA_y)^2}{2m} + \frac{eg\hbar}{4m}\hat{\sigma}.B + \frac{\hbar e}{4m^2c^2}\hat{\sigma}.\left[(\hat{p} + eA^B) \times \hat{E}\right], \quad (1)$$

where $\hat{\sigma}$ is the vector of Pauli spin matrices, $A^B = (0, A_y, 0)$ is the Landau gauge associated with the external $B$ field, and $g\hbar/4m$ is the Zeeman splitting strength. Eq. (1) can be transformed to:

$$H = \frac{1}{2m}\left(p_x + \left[\frac{\hbar e}{4mc^2}\sigma_i E_j \varepsilon_{ijx}\right]\right)^2 + \frac{1}{2m}\left(p_y + e\left[A_y + \frac{\hbar}{4mc^2}\sigma_i E_j \varepsilon_{ijy}\right]\right)^2 + \frac{eg\hbar}{4m}\sigma.\vec{B}, \quad (2)$$

after ignoring the higher order terms. Performing a unitary transformation of Eq. (2) with the rotation operator $U$, so that all local $B$ field orientations are transformed to the $z$-axis, we obtain

$$H' = UHU^+ = \sum_{k=x,y,z} \frac{1}{2m}\left(p_k + e\left[A_k + \frac{\hbar}{4mc^2}U\sigma_i E_j \varepsilon_{ijk}U^+ + \frac{\hbar}{e}U\partial_k U^+\right]\right)^2 + \frac{eg\hbar}{4m}\sigma_z B_z. \quad (3)$$

Under the adiabatic condition, the 2-by-2 matrix $U\partial_k U^+$ becomes diagonal, hereafter represented by matrix $A^M$ (whose components are the monopole potential). Taking note that $E$ field is vertical to the 2DEG plane in the Rashba system, the transformation $U(\sigma_i E_j \varepsilon_{ijk})U^+$ is equivalent to rotating the laboratory $z$-axis to the $B$ field axis (see Fig. 2). Thus, the gauge fields comprise of the $A^{SU(2)}$ term from the spin-orbital effects, and the topological term arising from the net chirality of the $B$ field, $A^M = \sigma_z |A_{ij}|\delta_{ij}$, where $i, j$ are respectively the row, column index of a 2-by-2 matrix. In the presence of both U(1)⊗U(1) and SU(2) gauge fields, as well as scattering with electron, phonon, or impurities, the electron wave-vector changes randomly as it drifts from source to drain. The random motion would then propagate an electron back to its initial spin state through various closed paths before it exits the device. As each closed path is associated with a Berry phase solid angle, an electron would acquire a phase equal to the solid angle due to the $B$ field's chirality, if the motion is sufficiently random. The diagonal component of the monopole gauge matrix can be found by the following path-integral method [22,23] for spin parallel to field,

$$|\psi(x_{ini},t_f)\rangle = v(t,t_0)|\psi(x_{ini},t_0)\rangle = \int \prod_{i=0}^{N} \prod_{j=1}^{N} v(t_{i+1},t_i)dx_j |\psi(x_{ini},t_0)\rangle, \qquad (4)$$

where $v(t,t_0)$ is the evolution operator between times $t_0$ and $t$. The final spin state $|x_{ini},t_f\rangle$ differs from the initial spin state $|x_{ini},t_0\rangle$ only by a phase angle:

$$\langle x_{ini}|\psi(x_{ini},t_f)\rangle = \langle x_{ini}|\int \prod_{i=0}^{N} \prod_{j=1}^{N} v(t_{i+1},t_i)dx_j |\psi(x_{ini},t_0)\rangle = \int \prod_{j=1}^{N} e^{iS/\hbar} dx_j. \qquad (5)$$

Therefore, $\langle x_{ini}|\psi(x_{ini},t_f)\rangle = e^{iS/\hbar}$, and we can define an evolution operator $U$ in a general form, i.e. $|\psi(x,t)\rangle = e^{iS/\hbar}|\psi(x_{ini},t_0)\rangle$. We showed that in a dynamic spinor system, $S$ corresponds to the action $S[n(t)] = \hbar \int_0^t iz^+ \partial_t z dt$, where $n(t)$ refers to the spinor vector at time $t$ [22]. An expansion of the action leads to:

$$\frac{S}{\hbar} = \int_0^t iz^+ \partial_t z dt = \int_0^t \left(\frac{1-\cos\theta}{2}\right)\partial_t \phi dt = \int \frac{(1-\cos\theta)}{2} \widehat{\nabla}_r \phi . d\widehat{r} . \qquad (6)$$

Considering the evolution over a short time interval $\Delta t$, the above integration leads to $A = \hbar/e(U\partial_r U^+) = \hbar/e(\partial_r[S/\hbar])$ or

$$\int \widehat{A}.d\widehat{r} = \pm \int \frac{\hbar}{2e}(1-\cos\theta)\widehat{\nabla}_r \phi . d\widehat{r} , \qquad (7)$$

for spin parallel (+) and anti-parallel (−) to field, respectively. The effect of the gauge fields on the electron motion can be studied by examining the individual effect of $A^{SU(2)}$ and $A^M$. The partial spin polarization induced by the external $B$-field can be described by $|\psi\rangle\langle\psi| = c^\uparrow|\psi_\uparrow\rangle\langle\psi_\uparrow| + c^\downarrow|\psi_\downarrow\rangle\langle\psi_\downarrow|$, where $c^\uparrow$ and $c^\downarrow$ are some function of temperature or other material parameters, and with $c^\uparrow + c^\downarrow = 1$. In our system, a weak spin polarization has been assumed so that $c^\uparrow \approx c^\downarrow = 0.5$. As $A^B$ acts on both up and down spin in the same transverse direction, a weak vertical polarization implies that $A^B$ would not contribute to the spin Hall effect. We will thus focus on the effective magnetic fields generally prescribed by the electromagnetic field tensor of

$$F_{\mu\nu} = \partial_\mu A_\nu - \partial_\nu A_\mu + \frac{ie}{\hbar}[A_\nu, A_\mu]. \tag{8}$$

Focusing on the spin-dependent part of the curvature, and using the relation $[U\lambda_\nu U^+, U\lambda_\mu U^+] = U[\lambda_\nu, \lambda_\mu]U^+$ where $\lambda_{\nu,\mu}$ is an arbitrary vector component, an effective field for the SU(2) gauge can be obtained:

$$\widehat{B}_z = \frac{\hbar e}{8m^2 c^4}(U\widehat{\sigma}U^+ \cdot \widehat{E})E_z \widehat{n}_z, \tag{9}$$

where $\widehat{n}_z$ is a vertical unit vector. The non-abelian nature of the gauge arises from the non-commutativity of the SU(2) spin algebra. Noting that $\widehat{E}$ is vertical in the 2DEG system, we find that an electron traveling in the lab x-axis with spin parallel to B field axis will experience a transverse force of

$$|\widehat{F}_y| = (\hbar e^2/8m^2 c^4)(\sigma_z \cos\theta) v_x E_z^2. \tag{10}$$

Note that the local B field configuration has a net chirality with a small solid angle $\Omega$, hence the angle $\theta$. The approximation $U(\sigma_i E_j \varepsilon_{ijk})U^+ = (E_j U \sigma_i U^+)\varepsilon_{ijk} \approx (\sigma_i E_j)\varepsilon_{ijk}$ which holds for small $\theta$ would lead to a force $|\widehat{F}_y| = (\hbar e^2/8m^2 c^4)\sigma_z v_x E_z^2$, which is consistent with Eq. (10).

From Eq. (8), the effective magnetic field due to the U(1)$\otimes$U(1) gauge was obtained from the curvature of the monopole gauge, which is abelian:

$$\widehat{B}_z = \frac{\hbar}{2eR^2}\sigma_z \widehat{n}_z, \tag{11}$$

where R is the monopole radius (see Fig. 2b). The force related to this field is of Lorentz-type, $|\widehat{F}_y| = \frac{\hbar}{2R^2}\sigma_z v_x$. Inspection of the transverse force expressions derived show that both are spin-

dependent and point in opposite directions for the two spin orientations. In addition, the effective magnetic field directions suggest that the two forces can be designed to reinforce or to oppose one another, thus enhancing or cancelling the spin Hall effect. We consider a device where the $B$ and $E$ fields are along the $-z$ and $+z$ directions, respectively, and their magnitudes are such that the resultant spin Hall effects completely cancel one another. As one modulates the asymmetric $E$ field by varying the gate potential [5,6], the transverse $F^{SU(2)}$ force from the SO interactions declines in strength, resulting in an increase of the spin Hall current. The device can thus be used to turn on and off the spin Hall effect.

To illustrate the effects in a practical system, we consider a GaAs 2DEG with material parameters: effective electron mass $m^*=0.067m_0$, charge density $n_e=10^{11}$ cm$^{-2}$, Fermi energy $E_F=3.55$ meV, Fermi velocity $v_F=1.36 \times 10^5$ ms$^{-1}$. The external $B$ field is applied on the 2DEG plane such that it inscribes a circular distribution of radius $r=5$ nm, with a field orientation $\theta$ from the vertical, and a radially outward azimuthal direction (see Fig. 2b). We consider a non-equilibrium condition due to the application of an $E_x$ field. The typical Rashba constant $\alpha=4 \times 10^{-12}$ eVm in a GaAs 2DEG translates to an approximate in-built $E$ field strength $E_z=4.8 \times 10^{11}$ Vm$^{-1}$. The application of a gate voltage can affect the band-bending within the 2DEG heterostructure, thus modifying the value of $\alpha$ and hence $E_z$ of up to ~50% [5,6]. Adjusting the gate voltage to vary the $E_z$ field, we found (Fig. 3) that the total average transverse force can be modulated to switch on and off the spin Hall current. We illustrated this for three $B$ field orientations of $\theta = 1°, 1.5°, 2°$ (see Fig. 2 for $\theta$). In each case the $E$ field dependent SU(2) force induced by spin-orbit coupling is tuned in magnitude against the constant (but opposite) topological force. At a particular value of $E$ field the total transverse force is zero, thus achieving a complete cancellation of the spin Hall effect.

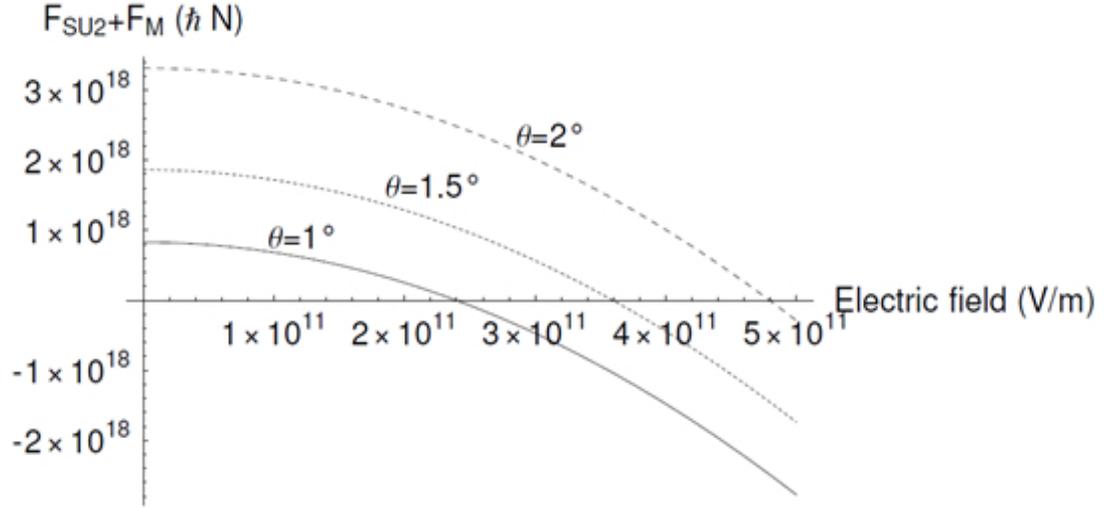

FIG. 3. Numerical evaluation of the two forces in opposing directions, for three values of the field angle $\theta$ (deg). As the electric field is increased, the component of the force due to spin-orbit coupling becomes significant compared to the constant topological force. At a critical $E$ field value, the two forces cancel one another completely, switching off the spin Hall effect.

In conclusion, we have designed a spin Hall system that is easy to realize and detect. We analyzed the forces acting on a spin-polarized current in a 2DEG heterostructure device in the presence of external vertical $B$ fields with net chirality, and Rashba SO interactions within the 2DEG. The spin up electrons will experience a transverse force due to the non-abelian nature of the SU(2) gauge arising from the SO effects and the topological gauge fields arising from the chirality of the external $B$ fields. The advantage of the device is that the essentially perpendicular external $B$ field induces a vertically-aligned spin current which can experience the full effects of the SU(2) and U(1)⊗U(1) gauge fields. This removes the need for very short and clean device, simplifying experimental effort to implement. Additionally, the tunability of the magnitude and direction of the SU(2) gauge field enables either a gate-voltage induced enhancement or cancellation of the spin Hall effect arising from the constant topological field, thus enabling the device to turn "on" and "off" the spin Hall effect. It is worth noting that the tunability of the SU(2) gauge is based on the physics of intrinsic but not the extrinsic spin Hall. This device could thus be used to determine the type of spin Hall effect which is predominant in the 2DEG system.

**ACKNOWLEDGEMENT**

We would like to thank S.Q. Shen for discussion. We thank the National University of Singapore (NUS) and the Data Storage Institute for funding theoretical research in spintronics and magnetic physics. X.-J.L would like to thank the NUS for supporting spintronic research under the NUS academic research, Grant No. WBS: R-144-000-172-101.